# Reorganization of functionally connected brain subnetworks in high-functioning autism

Enrico Glerean* (a), Raj Kumar Pan (a), Juha Salmi (a,b), Rainer Kujala (a), Juha Lahnakoski (a), Ulrika Roine (a), Lauri Nummenmaa (a,c), Sami Leppämäki (d,e), Taina Nieminen-von Wendt (f), Pekka Tani (e), Jari Saramäki (a), Mikko Sams (a), Iiro P. Jääskeläinen (a,g)

a) Department of Neuroscience and Biomedical Engineering (NBE), Aalto University, Espoo, Finland
b) Institute of Behavioural Sciences, Division of Cognitive and Neuropsychology, University of Helsinki, Finland
c) Turku PET Centre and Department of Psychology, University of Turku, Turku Finland
d) Finnish Institute of Occupational Health, Helsinki, Finland
e) Department of Psychiatry, Helsinki University Central Hospital, Helsinki, Finland
f) Neuropsychiatric Rehabilitation and Medical Centre Neuromental, Finland
g) Advanced Magnetic Imaging (AMI) Centre, Aalto University, Espoo, Finland
* Corresponding author: Enrico Glerean
Department of Neuroscience and Biomedical Engineering, School of Science, Aalto University; P.O. Box 12200, FI-00076 AALTO, Finland;
phone: +358 40 1877599; email: enrico.glerean@aalto.fi

Keywords: Autism Spectrum Disorder; Functional Connectivity; fMRI; Graph Theory; Intersubject Similarity; Network Modularity

Running title: Reorganization of functional subnetworks in autism

# Abstract

## Background

Previous functional connectivity studies have found both hypo- and hyper-connectivity in brains of individuals having autism spectrum disorder (ASD). Here we studied abnormalities in functional brain subnetworks in high-functioning individuals with ASD during free viewing of a movie containing social cues and interactions.

## Methods

Thirteen subjects with ASD and 13 matched-pair controls watched a 68 minutes movie during functional magnetic resonance imaging. For each subject, we computed Pearson's correlation between haemodynamic time-courses of each pair of 6-mm isotropic voxels. From the whole-brain functional networks, we derived individual and group-level subnetworks using graph theory. Scaled inclusivity was then calculated between all subject pairs to estimate intersubject similarity of connectivity structure of each subnetwork. Additional 27 individuals with ASD from the ABIDE resting-state database were included to test the reproducibility of the results.

## Results

Between-group differences were observed in the composition of default-mode and a ventro-temporal-limbic (VTL) subnetwork. The VTL subnetwork included amygdala, striatum, thalamus, parahippocampal, fusiform, and inferior temporal gyri. Further, VTL subnetwork similarity between subject pairs correlated significantly with similarity of symptom gravity measured with autism quotient. This correlation was observed also within the controls, and in the reproducibility dataset with ADI-R and ADOS scores.

## Conclusions

Reorganization of functional subnetworks in individuals with ASD clarifies the mixture of hypo- and hyper-connectivity findings. Importantly, only the functional organization of the VTL subnetwork emerges as a marker of inter-individual similarities that co-vary with behavioral measures across all participants. These findings suggest a pivotal role of ventro-temporal and limbic systems in autism.

# Introduction

Social and communication disturbances, and restricted or repetitive behavior constitute the core symptoms in high-functioning individuals with autism spectrum disorder (ASD). Reduced ability in, for example, perceiving subtle social cues and understanding the intentions of others make social interactions and forming of social relationships challenging to these individuals. Imaging and genetic studies characterize ASD as a manifestation of subtle abnormalities in brain connectivity in affected individuals (1). In functional imaging, what has been reported is a mix of findings from reduced connectivity (hypoconnectivity) to increased connectivity (hyperconnectivity; see (2) for a recent review). These findings vary according to population under investigation (developing *vs*. adults) and scanning paradigm (multiple types of active tasks or resting state).

Hypoconnectivity has been observed between prefrontal and posterior brain areas (3–6) (see (7) for a review), between other areas implicated in social cognition (8–11), as well as between subcortical and cortical structures in the sensory and motor systems (12; 13). A recent large scale study (14) suggested short and long distance hypoconnectivity across the whole ASD brain, with the exception of hyperconnectivity between subcortical and cortical structures. In other studies, hyperconnectivity has been observed locally in occipital (15), frontal, and temporal areas (16; 17), as well as in amygdala (18). Hyperconnectivity has also been reported in large-scale cortico-cortical (19–21), and cortico-subcortical networks (22; 23). Majority of hyperconnectivity observations have been in children or adolescents with ASD. In adults, hyperconnectivity has been reported between posterior cingulate, temporal lobe, and parahippocampal gyrus (6) as well as between amygdala and medial prefrontal cortex (9).

Graph-theoretical tools have been increasingly used in the analysis of functional brain connectivity (24). In such analysis, functional brain networks are described as consisting of nodes corresponding to voxels or regions of interest. These nodes are connected by links representing functional relationships as inferred from correlations of the functional activity time series of each pair of nodes. Patterns of functional relationships can then be described at several levels, from the properties of individual nodes and links (micro-level) to features of the global network (macro-level), and the intermediate (mesoscopic) level of subnetworks, also known as subgraphs, modules or communities (24; 25). Similarly as in (26), we adopted the term "subnetwork" to stress the graph-theoretical aspect of our approach.

The overall organization of a network's links typically reflects its function. This functional organization may not be visible at the micro level of individual nodes and links, or at the macro level of network summary statistics. Rather, it is apparent at the mesoscopic level of groups of densely interlinked nodes – subnetworks – that can be inferred from network structure (27). Most ASD functional connectivity research has focused on link or node-level differences. There are few graph theoretical ASD studies that have adopted the mesoscopic approach, either in structural imaging (28) or in combined structural and functional imaging at rest (29). However, whole-brain voxel-wise comparison of subnetworks of individuals with ASD and control subjects has not been realized to date using graph-theoretical tools.

Finally, functional brain connectivity abnormalities in ASD have been studied to date with subjects either not performing any task (*i.e.*, "resting state") or relatively simple tasks targeting to activate specific brain networks. At behavioral level, however, movie clips depicting various social cues and interactions appear to be more effective than isolated perceptual-cognitive tasks in capturing the complex and individualistic autistic traits (30; 31). Thus, it can be hypothesized that observing naturalistic social and emotional stimuli such as movies reveal the underlying functional connectivity abnormalities more closely related to everyday social interaction than resting state or the focused tasks designed to engage specific cognitive functions (32). Supporting this view, deviations in brain function in autistic individuals have been recently characterized during free viewing of movies (33–35). However, there are currently no reports on possible abnormalities in the configuration of the large-scale functional brain-network topography in ASD during free viewing of dynamic social interactions.

Here, we specifically hypothesized that the previously reported mixed hypo- and hyper-connectivity is reflected as differences in the composition of functional subnetworks in ASD and control subjects. Furthermore, we hypothesized that such differences co-vary with autistic symptom severity. To specifically study alterations in functional subnetworks in subjects with ASD during social cognition, we analyzed our previously published dataset (34) where 13 high-functioning autistic and 13 matched-pair control subjects' brain hemodynamic activity was measured with fMRI during free viewing of a drama movie containing social cues and interactions. Furthermore, to test the reproducibility of our findings, we also considered 27 high functioning individuals with ASD from the ABIDE resting state dataset (14).

# Methods and Materials

## Participants

The participants were 13 high functioning individuals with ASD (mean age 29 years, s.e.m. 1.7 years, range 20–41 years, all males) and 13 healthy male controls (mean age 28 years, s.e.m. 2.1 years, range 19–47 years) – from now on labeled as "neurotypical" (NT) – matched for age and IQ (see (34) for details). The participants with ASD filled the criteria for Asperger syndrome based on ICD-10 criteria. To quantify where subjects of the current study were on the autistic continuum, Autism quotient (AQ) (36), translated into Finnish (37), was obtained from all participants. AQ significantly differed ($p < 10^{-5}$) between the groups with ranges 6–35 (NT), 17–43 (ASD) and mean values of 12.5 (2.1 s.e.m.) for NT and 30.5 (2.1 s.e.m) for ASD.

## Stimulus

The stimulus was the Finnish feature film "The Match Factory Girl" (Aki Kaurismäki, 1990, length: 68 minutes).

## MRI data acquisition and preprocessing

MR imaging was performed with a Signa VH/i 3.0T scanner using a quadrature 8-channel head coil. A total of 1980 functional volumes were obtained, each consisting of 29 gradient-echo planar axial slices (thickness 4 mm, 1 mm gap between 5 slices, TE 32ms, TR 2000ms, see (34) for details).

Preprocessing was performed with FSL using the FEAT pipeline: removal of first 29 volumes (corresponding to movie titles), motion correction, 6 mm spatial smoothing, two steps co-registration to MNI 152 2mm template, temporal filtering at 0.01–0.08 Hz. Data were then spatially downsampled to 6 mm isotropic voxels resulting in 5184 brain grey matter voxels. To control for head motion confounds, motion parameters were regressed out. As all subjects had >95% volumes under framewise displacement threshold of 0.5 mm (38) and we failed to see any group differences in mean framewise displacement, we retained all timepoints for the analysis. However, we used the individual mean framewise displacement values as a nuisance variable for group level regression analysis (39). Mean brain signal was not regressed out since doing this reduces task effects (40) and can systematically bias functional network comparisons (41).

## Network Construction

For each subject we calculated Pearson's correlation between each pair of 6-mm isotropic voxels (nodes) time series, resulting in a 5184 x 5184 adjacency matrix with 13 434 336 correlation coefficients (weighted links). A sparse network was obtained by constructing minimum spanning tree (MST) followed by thresholding as in (25; 42). Network threshold was optimized as follows. For each link density from 0.1% (strongest 13 434 links, equivalent to retaining links with average r > 0.88) to 100% (all ~13.4 million links) the overlap of the resulting binarized networks was calculated between subject pairs (Figure S1). Given two graphs thresholded at density n% (i.e. retaining the strongest n% of all links), their overlap is given by the number of common links divided by the total number of links at that density. We selected the 2% density to focus on the maximally different networks constituted by the edges with highest correlation (r > ~0.69), as this has been suggested to provide most detailed parcellation of brain networks (26). This corresponds with previously accepted criteria (25).

## Computation of individual-subject functional subnetworks

We defined each functional subnetwork as a subset of nodes having a higher density of connections than expected on the average. We used the "Louvain method" (43), which maximizes the modularity of the detected partitions (44). We performed 100 optimization runs for each subject and selected the partitions that gave the highest value of modularity.

## Group consensus of subnetworks

To determine differences in the structure of subnetworks between ASD and NT, we first calculated a consensus partition for each group using a meta-clustering algorithm based on clustering clusters (45). Specifically, the set of partitions for each subject is transformed to a hyper-graph with each subnetwork representing a hyper-edge, i.e. an edge that can connect any number of vertices. Related hyper-edges are then grouped and collapsed together using METIS (46). The reduced number of hyper-edges was set as equal to the maximum number of subnetworks in any of the subject's partitions. Each node is then assigned to the collapsed hyper-edge where it participates most strongly. Finally, we matched the two groups consensus clustering labels with the Hungarian algorithm (47). The partition labels of individual subjects were also matched with the consensus subnetwork of their group. We then measured the group consistency of each node by counting the fraction of subjects for which the node belonged to the same subnetwork. This reflects the extent of agreement about the subnetwork label of the node. Our procedure is quite similar to that described by Alexander-Bloch and coworkers (25), however it is more general as it uses consensus partitions rather than the single most representative subject in the population.

## Labelling of functional subnetworks

Labels of subnetworks were assigned by firstly computing spatial overlap with known major subnetworks computed for a large number of subjects as reported in (48) and (26). Spatial overlap is defined as the Pearson's correlation between the spatial maps as in (49). Values and details are reported in Table S1. Finally, subnetwork labels were chosen manually and, when possible, matched with the quantitative results from Table S1. Furthermore, nodes were also labelled automatically by matching each node with its corresponding Automatic Atlas Labeling (AAL) or Harvard Oxford (HO) label. We reported AAL labels for cerebral cortical areas and HO labels for subcortical and cerebellar areas.

## Intersubject similarity of subnetworks

We estimated intersubject similarity for each subnetwork using Scaled Inclusivity (SI), (50–52). SI is a similarity measure defined for a subject pair and for a single node, as the intersection of the subnetworks to which the node belongs, normalized by the size of each of the two subnetworks. Since SI is computed for a single node, to consider the similarity across subjects at subnetwork level we first considered the NT group consensus subnetworks. Next, for a chosen subnetwork we computed the median SI of all the nodes in this subnetwork for each subject pair. This produced an intersubject similarity matrix across all NT subjects and individuals with ASD for the chosen subnetwork (see Figure 1 for a schematic). Finally, we obtained the intersubject similarity matrices for each of the NT group subnetworks, where each element describes the level of similarity for the specific subnetwork for the subject pair. Each similarity matrix was tested for group differences by computing difference of the means scaled by the variance (i.e. comparable to a t-value) for the within group similarity values. P-values were computed with permutation tests for all comparisons (1 million permutations). Effect size for each group comparison were calculated with the MES toolbox (53) and we

Glerean et al. (2015) Reorganization of functional subnetworks in autism reported values of Hedges' g. We also tested each similarity matrix with a model matrix based on the similarity of AQ scores with Mantel test with one million permutations, and the effect size reported is the correlation coefficient between the two matrices.

Similarity between individuals' AQ scores was computed by considering separately the five domains of the AQ score (social skills; communication skills; imagination; attention to detail; and attention switching/tolerance of change), so that each subject was characterized with a five-dimensional vector of AQ subscale scores. Similarity of AQ scores (a real value between 0 and 1) was then derived from the Euclidean distance between the individual vectors (see Lemma 8 in (54)). To control for head motion, we also computed a intersubject similarity matrix based on mean framewise displacement as a measure of intersubject similarity of average head motion. Code used in this paper is available at https://github.com/eglerean/hfASDmodules.

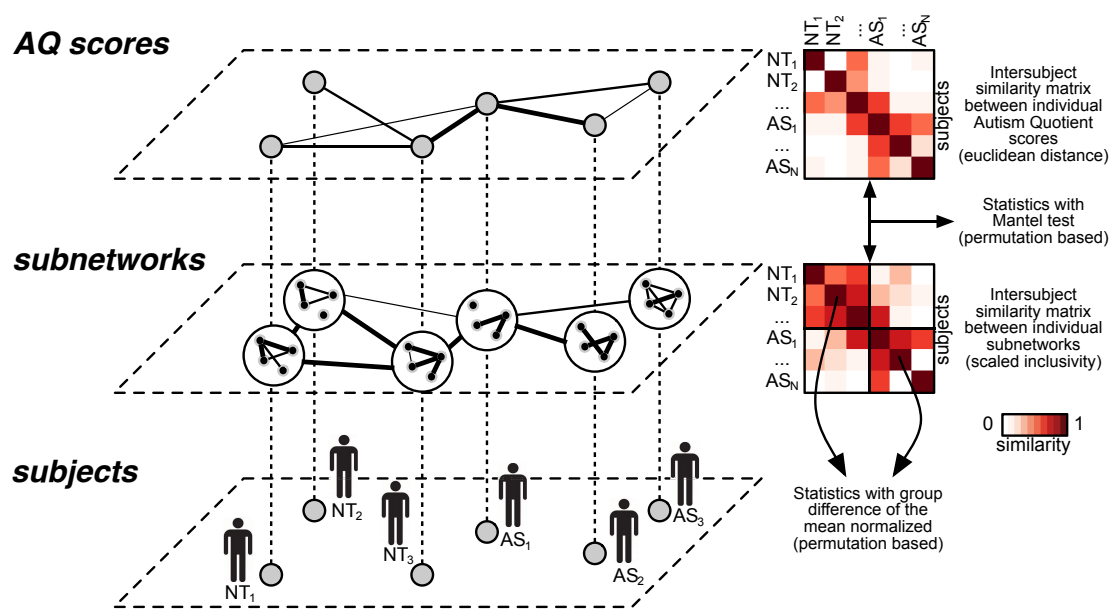

Figure 1 – A schematic representation of the intersubject analysis framework. For two groups of subjects (bottom layer), we can compute the similarity between each subject pair by using functional brain data at the level of subnetworks (middle layer) or behavioral scores (top layer). These layers are described as networks using adjacency matrices also known in this case as intersubject similarity matrices. Two types of statistical tests can then be run: a group difference within a layer, in which the within groups values of the adjacency matrix are compared (bottom adjacency matrix, where the group comparison tests whether the within NT group similarity is higher than the within ASD group similarity). The second test is the so-called Mantel test, in which the two adjacency matrices are compared with each other by correlating the corresponding values of the top off-diagonal triangle. In the latter case, also the between group similarity values are used making the Mantel approach more strict.

## Individual microscopic network properties versus Autism Quotient score

We correlated micro-level properties of nodes and links with individual AQ scores using Spearman correlation. Specifically, we computed the node strength as the sum of the weights of the links connected to a node. The significance threshold was computed by permuting the subjects' labels (100 000 permutations). We took the $95^{th}$ percentile of the max-statistics (for negative values, it is the $5^{th}$ percentile of the min-statistics) to correct for multiple comparisons as described in (55) which yielded correlation thresholds of -0.480 and 0.459. We then considered all the links in the individual networks with the 2% link density (~0.3 million links). We computed the Spearman correlation between individual AQ and link weights. To control for multiple comparisons, we used the false discovery rate cluster-based statistics as described in (56), which is an extension of the Network-based Statistics method (57). This yielded a threshold of 0.695 for positive AQ-link weight correlations and of -0.650 for negative correlations for clusters at a corrected $p < 0.05$ significance. For completeness we also computed macro-level network properties: mean link weight, clustering, average path length and efficiency (24).

## Reproducibility dataset

To test the reproducibility of the proposed intersubject similarity of movie subnetworks predicted by the similarity of the symptoms severity, we selected 27 subjects from the ABIDE database. Although these subjects were scanned in the resting state, we considered the group consensus NT subnetworks from the movie watching as reference subnetworks. By considering the reference subnetworks identified when processing social content, we test whether the same subset of regions showed a similar disruption at subnetwork level also during rest. Furthermore, ABIDE subjects were diagnosed using Autism Diagnostic Interview Revised (ADI-R) (58) and Autism Diagnostic Observation Schedule (59). The reproducibility test would then assess the validity of our findings for a different scanning paradigm and for other diagnostic tools. Further details on ABIDE subjects selection, preprocessing and quality control are reported in supplemental materials.

# Results

Whole-brain functional connectivity analysis disclosed 12 subnetworks in the NT subjects that are depicted with different colors on cortical surface on the left-hand side of Figure 2. For a detailed display of each subnetwork see Figure S2 or see the fully browsable results on NeuroVault (60) at http://neurovault.org/collections/437/. These 12 subnetworks consisted of (from bottom to top) 1) Default-mode (DM), 2) Language (LAN), 3) Auditory (AUD), 4) Salience (SAL) 5) Parietal, 6) Dorsal attention (DA), 7) Sensorimotor (SM), 8) Visual primary (V1) 9) Ventro-temporal limbic (VTL) – comprising subcortical areas (amygdala, nucleus accumbens, putamen, caudate, thalamus) as well as the anterior part of the ventral visual pathway and part of ventro-medial prefrontal cortex 10) Precuneus 11) Cerebellum, and 12) Visual extrastriate (VIS). While there were no significant between-group differences at the macroscopic level tests of mean link weight, clustering, average path length and efficiency, the alluvial diagram in the middle part of Figure 2 shows that these functional subnetworks were reconfigured in subjects with ASD: A number of brain areas that constitute each subnetwork in NT subjects were shifted to other subnetworks in subjects with ASD. Significant group differences between median SI values of subject pairs were found in five of the twelve subnetworks: DM, AUD, DA, V1, and VTL (see Table 1).

The largest group difference was in the VTL subnetwork ($p = 8.402 \times e-10$, Hedges' $g = 1.041$). In ASD subjects the extent of VTL network was reduced so that thalamus, parts of the orbitofrontal cortex, and posterior medial-inferior temporal lobe structures were not consistently included in the VTL subnetwork. On the other hand, temporal poles (TPO) were included to a greater extent in the VTL subnetwork in ASD than in NT subjects. Furthermore, inferior temporal gyrus (ITG), parahippocampal gyrus (PHG) and right amygdala formed an independent subnetwork in ASD subjects disjoint from Putamen, Caudate and other subcortical areas in VTL (blue ASD subnetwork, Figure 2 right).

The DM subnetwork ($p = 2.041e-05$, Hedges' $g = 0.653$) revealed a reduction in the extent of fronto-medial and dorso-frontal areas belonging to default mode subnetwork in ASD subjects compared with NT controls (purple ASD subnetwork, Figure 2 right). In ASD these were included in a larger subnetwork together with the salience subnetwork (anterior cingulate gyrus – ACG –, anterior insula) and inferior frontal gyrus (IFG, yellow ASD subnetwork, Figure 2 right).

Glerean et al. (2015) Reorganization of functional subnetworks in autism

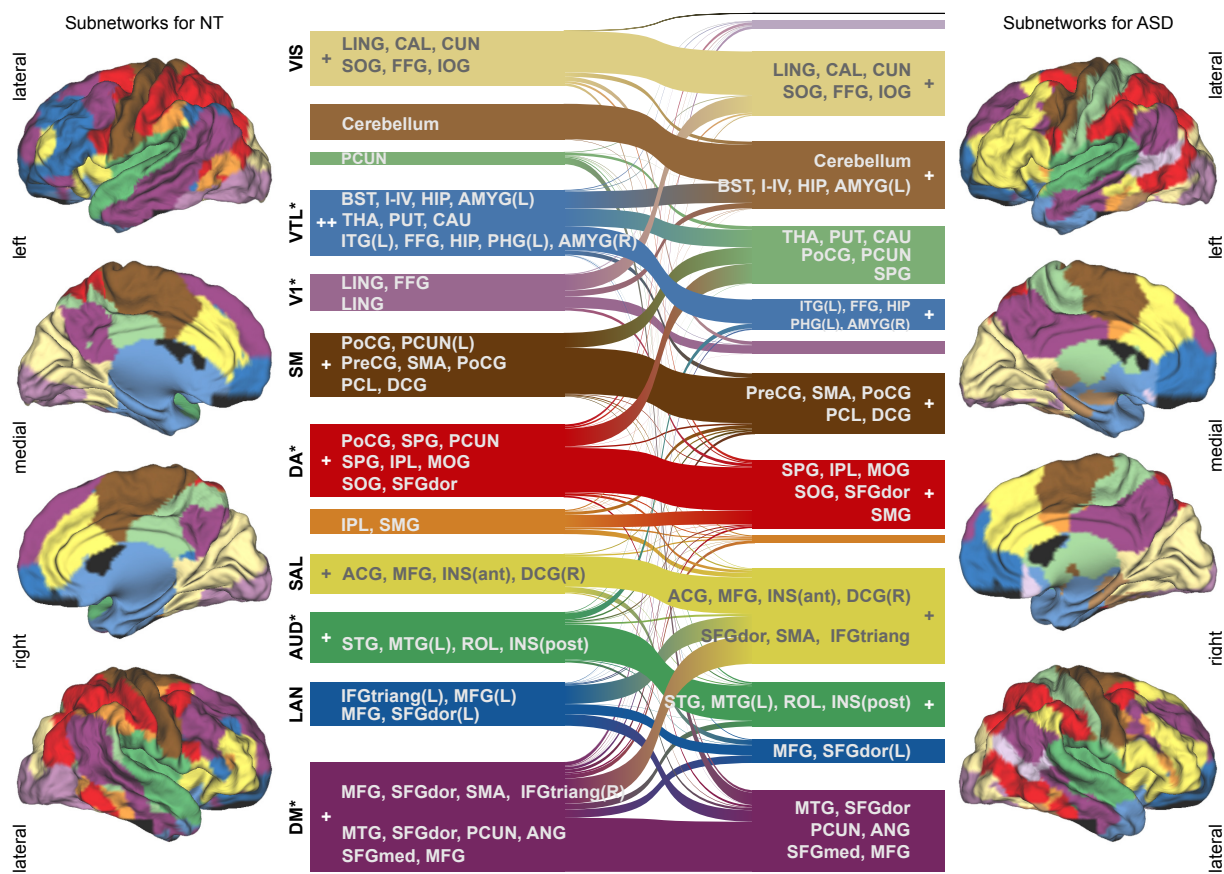

Figure 2 – Functional subnetwork similarities and differences between NT (left) and ASD (right) subjects. The subnetworks are color-coded and projected on lateral and medial surfaces of both hemispheres. The alluvial diagram in the middle uses the same color-coding. The height of each ribbon representing a subnetwork corresponds to the number of nodes that belong to the given subnetwork. Stars indicate statistically significant group difference: *significant at p<0.05, see also Table 1; plus signs indicate median consistency of all nodes within a subnetwork: +median subnetwork consistency>0.5, ++median subnetwork consistency > 0.75. Group consensus modules and consistency values for each node are available at http://neurovault.org/collections/437/. Ribbons with same color show related areas partitioned into similar subnetworks for both the groups. ACG: Anterior cingulum; AMYG: Amygdala; ANG: Angular gyrus; BST: Brainstem; CAL: Calcarine gyrus; CAU: Caudate; CUN: Cuneus; DCG: Middle cingulum; FFG: Fusiform gyrus; HES: Heschl gyrus; HIP: Hippocampus; IFGoperc: Opercular inferior frontal gyrus; IFGtriang: Triangular inferior frontal gyrus; INS: Insula; IOG: Inferior occipital gyrus; IPL: Inferior parietal lobule; ITG: Inferior temporal gyrus; LING: Lingual gyrus; MFG: Middle frontal gyrus; MOG: Middle occipital gyrus; MTG: Middle temporal gyrus; NAcc: Nucleus accumbens; OLF: Olfactory cortex; ORBinf: Orbital inferior frontal gyrus; ORBmid: Orbital middle frontal gyrus; ORBsupmed: Orbital medial frontal gyrus; ORBsup: Orbital superior frontal gyrus; PAL: Pallidum; PCG: Posterior cingulum; PCL: Paracentra lobule; PCL: Paracentral lobule; PCUN: Precuneus; PHG: Parahippocampal gyrus; PUT: Putamen; PoCG: Postcentral gyrus; PreCG: Precentral gyrus; REC: Gyrus rectus; ROL: Rolandic operculum; SFGdor: Superior frontal gyrus; SFGmed: Medial superior frontal gyrus; SMA: Supplementary motor area; SMG: Supramarginal gyrus; SOG: Superior occipital gyrus; SPG: Superior parietal lobule; STG: Superior temporal gyrus; THA: Thalamus; TPOmid: Temporal pole (middle); TPOsup: Temporal pole (superior).

Glerean et al. (2015) Reorganization of functional subnetworks in autism

| ID | Subnetwork name | Difference of the means normalized with variance (i.e. t-value) | P-value (via permutations) | Effect size for the difference of the means Hedges'g (95% c.i.) |
|---|---|---|---|---|
| 1 | Default mode (DM) | 4.126 | 2.041e-05* | 0.653 (0.377 0.922) |
| 2 | Language (LAN) | 0.789 | 0.218 | 0.204 (-0.135 0.524) |
| 3 | Auditory (AUD) | 4.057 | 3.759e-05* | 0.620 (0.306 0.962) |
| 4 | Salience (SAL) | -0.491 | 0.3124 | 0.348 (0.0317 0.671) |
| 5 | Parietal | -0.607 | 0.2725 | -0.024 (-0.33 0.285) |
| 6 | Dorsal attention (DA) | 3.014 | 0.001565* | 0.428 (0.117 0.763) |
| 7 | Sensorimotor (SM) | -0.171 | 0.4327 | -0.038 (-0.383 0.3) |
| 8 | Visual primary (V1) | 5.788 | 1.265e-09* | 0.537 (0.237 0.874) |
| 9 | Ventro-temporal limbic (VTL) | 10.112 | 8.402e-10* | 1.041 (0.709 1.42) |
| 10 | Precuneus | -1.322 | 0.09558 | -0.219 (-0.514 0.0893) |
| 11 | Cerebellum | 1.059 | 0.1457 | 0.068 (-0.233 0.363) |
| 12 | Visual exstrastriate (VIS) | 2.238 | 0.0137 | 0.165 (-0.162 0.52) |

Table 1 – The table reports group differences for each subnetwork as differences of the mean and effect size with confidence intervals. * = significant with Bonferroni correction at p<0.05;

### Intersubject similarity of subnetwork structure and the AQ score

We found a statistically significant relationship between intersubject similarity of subnetwork structure and AQ scores for the VTL subnetwork (Figure 3). Subject pairs with more similar AQ subscale scores had more similar VTL subnetwork SI ($r = 0.293$, $p = 0.000297$), independently of their diagnosis. When NT and ASD subjects were analyzed separately, the similarity was larger in NT subjects ($r = 0.549$, $p = 0.00922$) compared to ASD subjects ($r = 0.257$, $p = 0.0236$; test of the difference between correlation coefficients (61) NT > ASD $z = 2.168$, $p = 0.0301$, two tailed). When considering other subnetworks, we failed to see any other intersubject relationships between subnetwork structure and AQ. Intersubject similarity of average head motion did not correlate with AQ similarity and did not correlate with median SI.

Glerean et al. (2015) Reorganization of functional subnetworks in autism

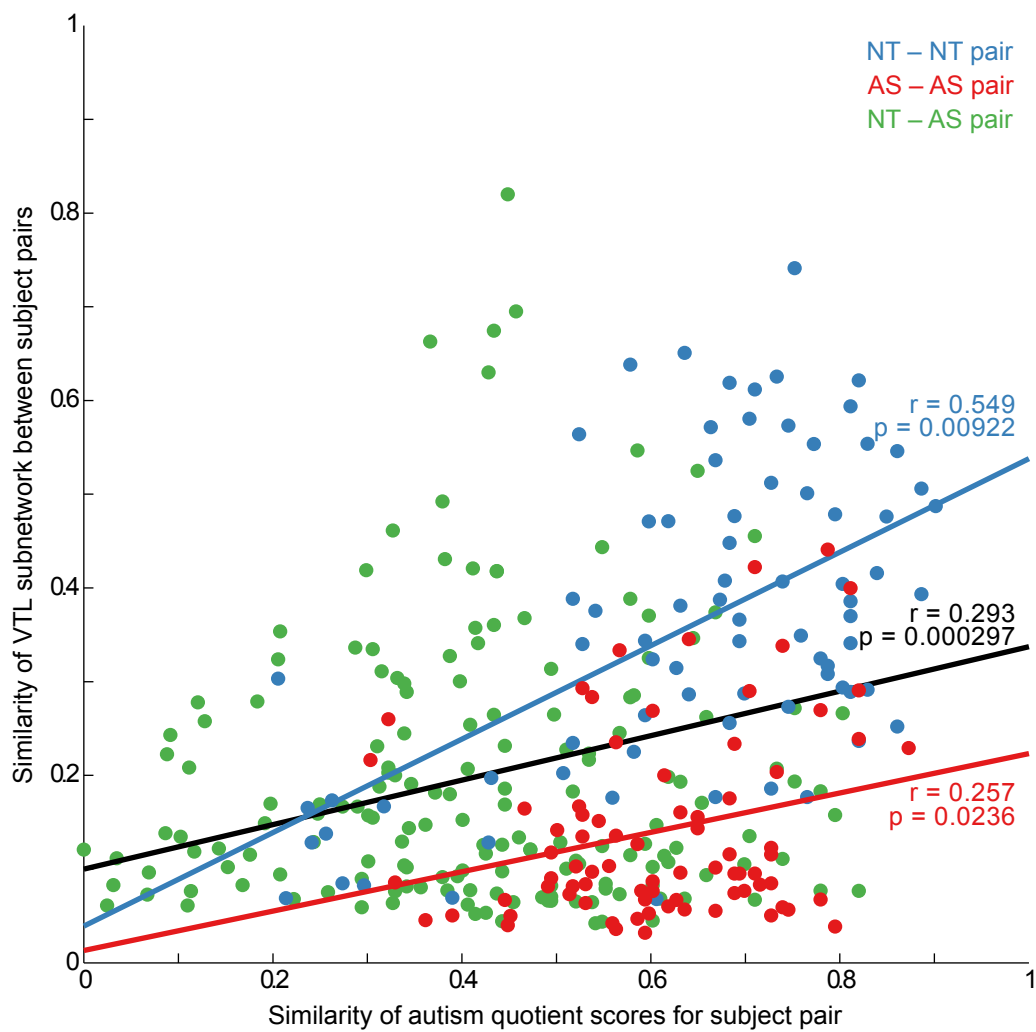

Figure 3 – Mantel test showing association between VTL subnetwork structure and autistic symptoms. Each dot is a pair of subjects showing their subnetwork similarity with median scaled inclusivity and behavioral similarity with AQ score vectors. Pairs are coded based on within groups (blue NT, red AS) and across groups (green). Mantel test results in the black interpolation line was performed using all data points. Mantel test results in blue (NT) and red (ASD) interpolation lines are only for within group values. Effect sizes are reported as correlation values and p-values were computed with permutations.

Node and link level results

Subjects with lower AQ scores exhibited significantly higher node strength in ACG and medial prefrontal cortex (MFG), dorsal part of the frontal gyrus, TPO, precuneus, and fusiform gyrus (FFG) (Figure 4A, peak coordinates in Table 2). Participants with higher AQ showed higher node strength in the posterior cingulate cortex (PCG), dorsal superior frontal gyrus (SFGdor), left IFG, and inferior parietal lobule (IPL). The connections (links) between areas that highly correlated with AQ score are reported as a summary connectivity matrix in Figure 4B, where nodes were grouped into anatomical regions (summary for all AAL regions in Figure S3). Low AQ was associated with higher functional connectivity across both long-distance (between frontal and parietal, frontal and occipital, as well as temporal and parietal) and within anatomical regions (i.e., the blue squares in the main diagonal of Figure 4B) such as the ACG, parahippocampal and superior parietal gyri. Few links were stronger for subjects with higher AQ, for example links within MFG and SFGdor or links between FFG and middle temporal areas.

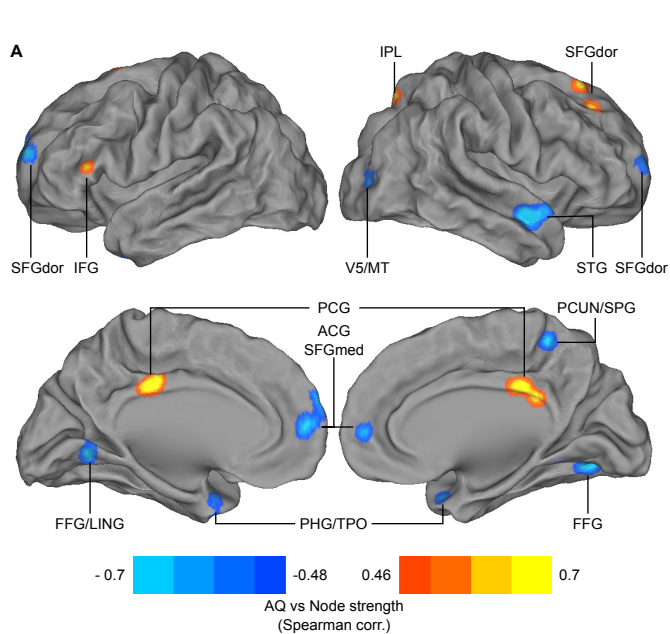

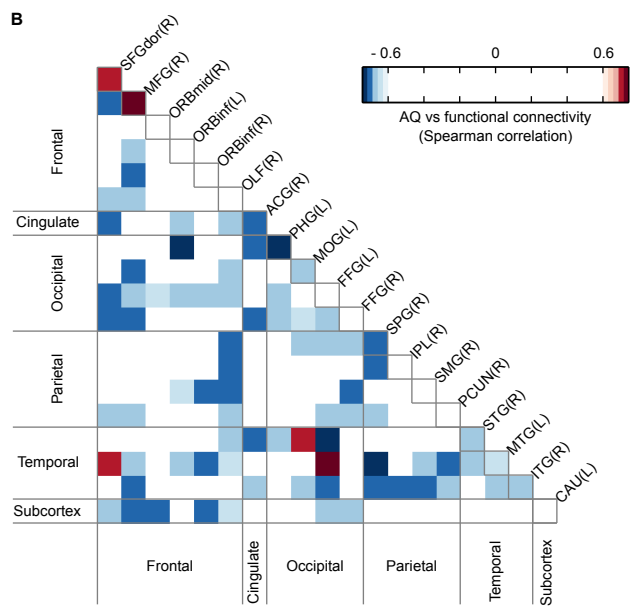

Figure 4. Node-level and link-level Spearman correlations with AQ scores. (A) Map of nodes whose strength values correlate with individual AQ scores. (B) Summary plot of link weight correlations with individual AQ scores. Only the strongest positively and negatively correlated links are reported (links in the 1st percentile). For a full summary connectivity matrix, see Figure S4. Each element of the pairwise connectivity matrix indicates the average of the correlations between AQ and link weights for all the links between a pair of anatomical regions. The main diagonal shows the average correlation for links within the respective region.

Glerean et al. (2015) Reorganization of functional subnetworks in autism

| Abbreviations (Fig.4) | MNI x | MNI y | MNI z | AQ vs Node strength (Spearman correlation) |
|---|---|---|---|---|
| ACG/SFGmed(L) | -10 | 50 | 14 | -0.63859 |
| ACG/SFGmed(L) | -10 | 56 | 26 | -0.61394 |
| STG | 50 | -4 | -10 | -0.59373 |
| PCUN/SPG | 8 | -46 | 56 | -0.59237 |
| SFGdor | -22 | 56 | 20 | -0.57764 |
| ACG/SFGmed | -10 | 56 | 14 | -0.5749 |
| SFGdor | 20 | 62 | 14 | -0.56566 |
| V5/MT | 32 | -82 | 8 | -0.54922 |
| ACG/SFGmed(L) | -10 | 56 | 20 | -0.54888 |
| TPO(R) | 32 | 8 | -28 | -0.53998 |
| FFG/LING(L) | -22 | -64 | -10 | -0.5345 |
| PHG/TPO(L) | -28 | 2 | **-30** | -0.52217 |
| FFG(R) | 26 | -70 | -10 | -0.51361 |
| ACG(R) | 8 | 50 | 8 | -0.49889 |
| IPL(R) | 14 | -70 | 50 | 0.4756 |
| IFG(L) | -52 | 26 | 14 | 0.50984 |
| SFGdor(R) | 20 | 32 | 44 | 0.51704 |
| PCG | 2 | -40 | 26 | 0.5773 |
| PCG | -4 | -34 | 32 | 0.60024 |

Table 2 – The table reports the voxels whose node strength correlates positively or negatively with AQ (Spearman correlation) and their Montreal Neurological Institute coordinates.

Reproducibility: subnetwork structure at rest and ADI-R/ADOS scores

When repeating the intersubject similarity analysis for the 27 ABIDE participants, we obtained similar results as those for the within ASD group i.e. only the VTL subnetwork showed significant correlation between subnetwork intersubject similarity and joint ADI-R/ADOS score similarity (p=0.0287) with moderate effect size (r = 0.17, Figure S4).

# Discussion

We studied with fMRI how functional subnetwork structure differs in individuals with ASD using novel graph theoretical tools applied to whole-brain functional networks without *a priori* assumptions on the nodes or links. We showed that ASD is characterized with significant reorganization of ventro-temporal-limbic and default-mode functional subnetworks. This reorganization is coupled to a mixture of micro-level hypo- and hyper-connectivity and the mesoscopic analysis clarifies the micro-level results. Moreover, the pattern of altered connectivity in the VTL subnetwork was associated with the degree of autistic symptoms, both in participants with ASD and controls. Altogether our findings suggest that aberrant organization of brain subnetworks may underlie social impairments in ASD.

Glerean et al. (2015) Reorganization of functional subnetworks in autism

We assumed that the drama movie drives functional network activity that cannot be easily seen during resting state, thus allowing us to elucidate functional network differences between NT and ASD subjects under conditions reflecting lifelike social environment. In the NT subjects, we observed subnetworks (Figures 2 and S2) closely resembling in composition those disclosed during resting state by other clustering methods (multidimensional clustering in (48), *infomap* graph clustering in (26), independent component analysis in (49)), supporting the validity of our analysis approach. There were, however, novel differences in the subnetworks of NT subjects, as compared with resting state studies. The VTL subnetwork, consisting of amygdala, hippocampus, parahippocampal cortex, fusiform gyrus, inferior temporal gyrus, temporal pole, thalamus, posterior aspects of orbitofrontal cortex, and striatum in the NT subjects is not consistently found in experiments using only a resting state condition (for a discussion (51)). Hence, we likely found the VTL subnetwork due to the use of a stimulus that engages social cognition and emotional processing in the subjects. This is also in line with studies showing different and more reliable brain connectivity in subcortical and limbic areas during task vs. resting-state conditions (62).

Notably, we observed a number of significant differences in the composition of subnetworks between ASD and NT subjects (Figure 2). In general, participants with ASD showed lower intersubject similarity of subnetwork structure (Table 1), likely reflecting the heterogeneity of the disorder (1) and idiosyncratic connectivity organization in ASD (63). The most robust group differences were observed in the VTL, DM and DA, as well as in subnetworks comprising visual and auditory areas.

Specifically, the coherent subnetwork activity between medial-frontal, inferior temporal, and subcortical structures is broken down in ASD subjects. Reconfiguration of the VTL subnetwork significantly correlated with severity of autistic symptoms as indexed by subject pairs with more similar VTL composition having similar AQ subscale scores (Figure 3). This important finding links the brain functional subnetwork-level differences to autistic symptoms, tentatively suggesting that the difficulties ASD individuals experience in social cognition are associated with an abnormal VTL subnetwork composition.

Notably, intersubject similarity of AQ score and VTL structure was also significant within the NT population (Figure 3). In the healthy population, high AQ has been reported to be associated with lower prosocial behavior (64) and with difficulties in voice processing (65). The connectivity between some of VTL areas is also known to be related to personality traits (66). When considering the reproducibility test of this finding with the ABIDE dataset, despite the differences of scanning paradigm and diagnostic tools, similarity of VTL was also correlated with ADI-R/ADOS intersubject similarity. This result points to a shared pattern between different autistic individuals in VTL subnetwork, independently of the presence or absence of stimulus. Although the complex stimulus was necessary to identify the VTL subnetwork, the resting-state result seems to be more related to the underlying structural connections.

While VTL subnetwork differences between NT and ASD subjects have not been previously investigated *per se*, these brain regions and their connectivity are known to be fundamental in ASD for a long time (67). Regions in VTL subnetworks are part of a larger distributed network involved in social cognition with previously reported hypo-connectivity in ASD involving FFG, amygdala, anterior hippocampus, insula, MFG, TPO, PCG, precuneus, Broca's area, the posterior superior temporal sulcus and temporo-parietal junction (11; 68). Impairment in the 'social motivation' circuit – amygdala, striatum and orbito-frontal cortex – has also been hypothesized to be a core feature of ASD (69). Specifically, in the striatum, caudate has been reported to be less connected with other subcortical areas in high functioning ASD adults than in NT ((13); see also (70)) and recent findings in genetic studies are showing how ASD genes have expression patterns highly specific to striatum (71). Our results also seem to point to the striatum role in ASD subnetwork reconfiguration, separating from amygdalas and other VTL areas (Figure 2). Finally, when considering the group differences in DM and DA composition, we corroborated previous observations obtained with other functional network analysis methods (5; 7; 20; 29).

In addition to the subnetwork-level differences, several micro-level (*i.e.*, node- and link-level) differences were observed within and between brain areas (Figure 4), consistent with previous studies showing, for example, higher node degree for ASD in SFGdor (14) and precuneus (72) as well as lower node degree for ASD in superior temporal gyrus and ACG (72). In addition, differences were observed in some nodes (left IFG, FFG , PHG) that are part of the VTL subnetwork in NT subjects. Importantly, the subnetwork-level analysis provides information that is not available at microscopic level. For example, the subnetwork-level analysis shows how the reduced connectivity of ACG (at node level inspection) in subjects with high AQ is due to ACG loosing its connections with the salience subnetwork and joining into a larger subnetwork involving SFGdor, Broca's area and middle frontal gyrus in ASD (see Figure 2). In a similar fashion, reduced connectivity in subcortical areas, FFG, and PHG does not simply mean disconnection in ASD: While a subset of VTL nodes in ASD isolates itself (ITG, FFG, light blue subnetwork figure 2 right hand side), striatum and thalamus form an anomalous subnetwork with PCG, precuneus and superior parietal cortex. Thus, simply looking at the connections each node has with other nodes (i.e., micro-level inspection) does not reveal the bigger picture of how the pattern of connectivity of that node is altered with respects to other nodes in the network, thus changing the functional role of the node – as well as the fine functional properties of the subnetworks that the node abandons and joins. To reveal and examine these effects and to resolve the micro-level mixed hypo- and hyperconnectivity findings, mesoscopic-level inspection of whole-brain network structure between NT and ASD subjects was needed.

In conclusion, our results suggest that mixture of hypo- and hyper-connectivity reported across previous ASD studies relates to specific differences in the composition of brain functional subnetworks. Anomalies in VTL subnetwork are associated with gravity of autistic symptoms, even within the NT group and even when considering resting state paradigm and different diagnostic tools. Using engaging stimuli and considering all intersubject variance of subnetworks might reveal consistent patterns also in other spectrum of disorders, like schizophrenia, that are characterized by high heterogeneity of symptoms and are related to neuronal connectivity dysfunction (73).

# Acknowledgments

The study was supported by the Academy of Finland (National Centres of Excellence program 2006–2011, grants #129670, #130412, #138145, #260054, #276643, #259752), Finnish Cultural Foundation (grant #150496) to JL, the aivoAALTO project grant from the Aalto University, Päivikki and Sakari Sohlberg Foundation, ERC Starting Grant #313000 to LN, Academy of Finland MIND program grant #265917 to LN. We thank Satu Saalasti for commenting the manuscript and Hanna Halme for preliminary connectivity analysis on the dataset. We acknowledge the computational resources provided by the Aalto Science-IT project.

# Financial Disclosures

The authors report no biomedical financial interests or potential conflicts of interest.

# References

1. Hernandez LM, Rudie JD, Green SA, Bookheimer S, Dapretto M (2014): Neural signatures of autism spectrum disorders: insights into brain network dynamics. *Neuropsychopharmacology*.

2. Maximo JO, Cadena EJ, Kana RK (2014): The implications of brain connectivity in the neuropsychology of autism. *Neuropsychol Rev* 24: 16–31.

3. Just MA, Cherkassky VL, Keller TA, Kana RK, Minshew NJ (2007): Functional and anatomical cortical underconnectivity in autism: evidence from an FMRI study of an executive function task and corpus callosum morphometry. *Cereb Cortex N Y N 1991* 17: 951–61.

4. Damarla SR, Keller TA, Kana RK, Cherkassky VL, Williams DL, Minshew NJ, Just MA (2010): Cortical underconnectivity coupled with preserved visuospatial cognition in autism: Evidence from an fMRI study of an embedded figures task. *Autism Res Off J Int Soc Autism Res* 3: 273–9.

5. Kennedy DP, Courchesne E (2008): The intrinsic functional organization of the brain is altered in autism. *NeuroImage* 39: 1877–85.

6. Monk CS, Peltier SJ, Wiggins JL, Weng S-J, Carrasco M, Risi S, Lord C (2009): Abnormalities of intrinsic functional connectivity in autism spectrum disorders. *NeuroImage* 47: 764–72.

7. Just MA, Keller TA, Malave VL, Kana RK, Varma S (2012): Autism as a neural systems disorder: A theory of frontal-posterior underconnectivity. *Neurosci Biobehav Rev* 36: 1292–1313.

8. Kleinhans NM, Richards T, Sterling L, Stegbauer KC, Mahurin R, Johnson LC, *et al.* (2008): Abnormal functional connectivity in autism spectrum disorders during face processing. *Brain J Neurol* 131: 1000–12.

9. Monk CS, Weng S-J, Wiggins JL, Kurapati N, Louro HMC, Carrasco M, *et al.* (2010): Neural circuitry of emotional face processing in autism spectrum disorders. *J Psychiatry Neurosci JPN* 35: 105–14.

10. Ebisch SJH, Gallese V, Willems RM, Mantini D, Groen WB, Romani GL, *et al.* (2011): Altered intrinsic functional connectivity of anterior and posterior insula regions in high-functioning participants with autism spectrum disorder. *Hum Brain Mapp* 32: 1013–28.

Glerean et al. (2015) Reorganization of functional subnetworks in autism

11. Gotts SJ, Simmons WK, Milbury LA, Wallace GL, Cox RW, Martin A (2012): Fractionation of social brain circuits in autism spectrum disorders. *Brain J Neurol* 135: 2711–25.
12. Villalobos ME, Mizuno A, Dahl BC, Kemmotsu N, Müller R-A (2005): Reduced functional connectivity between V1 and inferior frontal cortex associated with visuomotor performance in autism. *NeuroImage* 25: 916–25.
13. Turner KC, Frost L, Linsenbardt D, McIlroy JR, Müller R-A (2006): Atypically diffuse functional connectivity between caudate nuclei and cerebral cortex in autism. *Behav Brain Funct BBF* 2: 34.
14. Di Martino A, Yan C-G, Li Q, Denio E, Castellanos FX, Alaerts K, *et al.* (2014): The autism brain imaging data exchange: towards a large-scale evaluation of the intrinsic brain architecture in autism. *Mol Psychiatry* 19: 659–67.
15. Noonan SK, Haist F, Müller R-A (2009): Aberrant functional connectivity in autism: evidence from low-frequency BOLD signal fluctuations. *Brain Res* 1262: 48–63.
16. Shih P, Shen M, Ottl B, Keehn B, Gaffrey MS, Müller R-A (2010): Atypical network connectivity for imitation in autism spectrum disorder. *Neuropsychologia* 48: 2931–9.
17. Shih P, Keehn B, Oram JK, Leyden KM, Keown CL, Müller R-A (2011): Functional differentiation of posterior superior temporal sulcus in autism: a functional connectivity magnetic resonance imaging study. *Biol Psychiatry* 70: 270–7.
18. Murphy ER, Foss-Feig J, Kenworthy L, Gaillard WD, Vaidya CJ (2012): Atypical Functional Connectivity of the Amygdala in Childhood Autism Spectrum Disorders during Spontaneous Attention to Eye-Gaze. *Autism Res Treat* 2012: 652408.
19. Uddin LQ, Supekar K, Lynch CJ, Khouzam A, Phillips J, Feinstein C, *et al.* (2013): Salience network-based classification and prediction of symptom severity in children with autism. *JAMA Psychiatry* 70: 869–79.
20. Lynch CJ, Uddin LQ, Supekar K, Khouzam A, Phillips J, Menon V (2013): Default mode network in childhood autism: posteromedial cortex heterogeneity and relationship with social deficits. *Biol Psychiatry* 74: 212–9.
21. Supekar K, Uddin LQ, Khouzam A, Phillips J, Gaillard WD, Kenworthy LE, *et al.* (2013): Brain hyperconnectivity in children with autism and its links to social deficits. *Cell Rep* 5: 738–47.
22. Di Martino A, Kelly C, Grzadzinski R, Zuo X-N, Mennes M, Mairena MA, *et al.* (2011): Aberrant striatal functional connectivity in children with autism. *Biol Psychiatry* 69: 847–56.

23. Nair A, Treiber JM, Shukla DK, Shih P, Müller R-A (2013): Impaired thalamocortical connectivity in autism spectrum disorder: a study of functional and anatomical connectivity. *Brain J Neurol* 136: 1942–55.

24. Bullmore E, Sporns O (2009): Complex brain networks: graph theoretical analysis of structural and functional systems. *Nat Rev Neurosci* 10: 186–98.

25. Alexander-Bloch A, Lambiotte R, Roberts B, Giedd J, Gogtay N, Bullmore E (2012): The discovery of population differences in network community structure: New methods and applications to brain functional networks in schizophrenia. *NeuroImage* 59: 3889–3900.

26. Power JD, Cohen AL, Nelson SM, Wig GS, Barnes KA, Church JA, *et al.* (2011): Functional network organization of the human brain. *Neuron* 72: 665–78.

27. Fortunato S (2010): Community detection in graphs. *Phys Rep* 486: 75–174.

28. Shi F, Wang L, Peng Z, Wee C-Y, Shen D (2013): Altered modular organization of structural cortical networks in children with autism. (O. Sporns, editor.)*PloS One* 8: e63131.

29. Rudie JD, Shehzad Z, Hernandez LM, Colich NL, Bookheimer SY, Iacoboni M, Dapretto M (2012): Reduced functional integration and segregation of distributed neural systems underlying social and emotional information processing in autism spectrum disorders. *Cereb Cortex N Y N 1991* 22: 1025–37.

30. Klin A, Jones W, Schultz R, Volkmar F, Cohen D (2002): Visual fixation patterns during viewing of naturalistic social situations as predictors of social competence in individuals with autism. *Arch Gen Psychiatry* 59: 809–16.

31. Golan O, Baron-Cohen S, Hill JJ, Golan Y (2006): The "reading the mind in films" task: complex emotion recognition in adults with and without autism spectrum conditions. *Soc Neurosci* 1: 111–23.

32. Nummenmaa L, Saarimäki H, Glerean E, Gotsopoulos A, Jääskeläinen IP, Hari R, Sams M (2014): Emotional speech synchronizes brains across listeners and engages large-scale dynamic brain networks. *NeuroImage* 102 Pt 2: 498–509.

33. Hasson U, Avidan G, Gelbard H, Vallines I, Harel M, Minshew N, Behrmann M (2009): Shared and idiosyncratic cortical activation patterns in autism revealed under continuous real-life viewing conditions. *Autism Res Off J Int Soc Autism Res* 2: 220–31.

34. Salmi J, Roine U, Glerean E, Lahnakoski J, Nieminen-von Wendt T, Tani P, *et al.* (2013): The brains of high functioning autistic individuals do not synchronize with those of others. *NeuroImage Clin* 3: 489–97.
35. Pantelis PC, Byrge L, Tyszka JM, Adolphs R, Kennedy DP (2015): A specific hypoactivation of right temporo-parietal junction/posterior superior temporal sulcus in response to socially awkward situations in autism. *Soc Cogn Affect Neurosci*. doi: 10.1093/scan/nsv021.
36. Baron-Cohen S, Wheelwright S, Skinner R, Martin J, Clubley E (2001): The Autism-Spectrum Quotient (AQ): Evidence from Asperger Syndrome/High-Functioning Autism, Malesand Females, Scientists and Mathematicians. *J Autism Dev Disord* 31: 5–17.
37. Roine U, Roine T, Salmi J, Nieminen-Von Wendt T, Leppämäki S, Rintahaka P, *et al.* (2013): Increased Coherence of White Matter Fiber Tract Organization in Adults with Asperger Syndrome: A Diffusion Tensor Imaging Study. *Autism Res* 6: 642–650.
38. Power JD, Barnes KA, Snyder AZ, Schlaggar BL, Petersen SE (2012): Spurious but systematic correlations in functional connectivity MRI networks arise from subject motion. *NeuroImage* 59: 2142–54.
39. Yan C-G, Cheung B, Kelly C, Colcombe S, Craddock RC, Di Martino A, *et al.* (2013): A comprehensive assessment of regional variation in the impact of head micromovements on functional connectomics. *NeuroImage* 76: 183–201.
40. Van Dijk KRA, Hedden T, Venkataraman A, Evans KC, Lazar SW, Buckner RL (2010): Intrinsic functional connectivity as a tool for human connectomics: theory, properties, and optimization. *J Neurophysiol* 103: 297–321.
41. Gotts SJ, Saad ZS, Jo HJ, Wallace GL, Cox RW, Martin A (2013): The perils of global signal regression for group comparisons: a case study of Autism Spectrum Disorders. *Front Hum Neurosci* 7: 356.
42. Alexander-Bloch AF, Gogtay N, Meunier D, Birn R, Clasen L, Lalonde F, *et al.* (2010): Disrupted modularity and local connectivity of brain functional networks in childhood-onset schizophrenia. *Front Syst Neurosci* 4: 147.
43. Blondel VD, Guillaume J-L, Lambiotte R, Lefebvre E (2008): Fast unfolding of communities in large networks. *J Stat Mech Theory Exp* 2008: 6.
44. Newman M, Girvan M (2004): Finding and evaluating community structure in networks. *Phys Rev E* 69: 026113.

Glerean et al. (2015) Reorganization of functional subnetworks in autism45. Strehl A, Ghosh J (2003): Cluster Ensembles – A Knowledge Reuse Framework for. *J Mach Learn Res* 3: 583–617.

46. Karypis G, Kumar V (1998): *A Fast and High Quality Multilevel Scheme for Partitioning Irregular Graphs.*. (Vol. 20) . Retrieved from http://citeseerx.ist.psu.edu/viewdoc/summary?doi=10.1.1.106.4101.

47. Kuhn HW (1955): The Hungarian method for the assignment problem. *Nav Res Logist Q* 2: 83–97.

48. Thomas Yeo BT, Krienen FM, Sepulcre J, Sabuncu MR, Lashkari D, Hollinshead M, *et al.* (2011): The organization of the human cerebral cortex estimated by intrinsic functional connectivity. *J Neurophysiol* 106: 1125–1165.

49. Smith SM, Fox PT, Miller KL, Glahn DC, Fox PM, Mackay CE, *et al.* (2009): Correspondence of the brain's functional architecture during activation and rest. *Proc Natl Acad Sci U S A* 106: 13040–5.

50. Steen M, Hayasaka S, Joyce K, Laurienti P (2011): Assessing the consistency of community structure in complex networks. *Phys Rev E* 84: 016111.

51. Moussa MN, Steen MR, Laurienti PJ, Hayasaka S (2012): Consistency of Network Modules in Resting-State fMRI Connectome Data. (Y.-F. Zang, editor.)*PLoS ONE* 7: e44428.

52. Samu D, Seth AK, Nowotny T (2014): Influence of wiring cost on the large-scale architecture of human cortical connectivity. *PLoS Comput Biol* 10: e1003557.

53. Hentschke H, Stüttgen MC (2011): Computation of measures of effect size for neuroscience data sets. *Eur J Neurosci* 34: 1887–94.

54. Chen S, Ma B, Zhang K (2009): On the similarity metric and the distance metric. *Theor Comput Sci* 410: 2365–2376.

55. Nichols TE, Holmes AP (2002): Nonparametric permutation tests for functional neuroimaging: a primer with examples. *Hum Brain Mapp* 15: 1–25.

56. Han CE, Yoo SW, Seo SW, Na DL, Seong J-K (2013): Cluster-based statistics for brain connectivity in correlation with behavioral measures. *PloS One* 8: e72332.

57. Zalesky A, Fornito A, Bullmore ET (2010): Network-based statistic: identifying differences in brain networks. *NeuroImage* 53: 1197–207.

58. Lord C, Rutter M, Le Couteur A (1994): Autism Diagnostic Interview-Revised: a revised version of a diagnostic interview for caregivers of individuals with possible pervasive developmental disorders. *J Autism Dev Disord* 24: 659–685.
23

59. Lord C, Rutter M, Goode S, Heemsbergen J, Jordan H, Mawhood L, Schopler E (1989): Austism diagnostic observation schedule: A standardized observation of communicative and social behavior. *J Autism Dev Disord* 19: 185–212.

60. Gorgolewski KJ, Varoquaux G, Rivera G, Schwartz Y, Ghosh SS, Maumet C, *et al.* (2014): NeuroVault. org: A web-based repository for collecting and sharing unthresholded statistical maps of the human brain. *bioRxiv*010348.

61. Preacher KJ (2002): *Calculation for the test of the difference between two independent correlation coefficients [Computer software]*. . .

62. Mennes M, Kelly C, Colcombe S, Castellanos FX, Milham MP (2013): The extrinsic and intrinsic functional architectures of the human brain are not equivalent. *Cereb Cortex N Y N 1991* 23: 223–9.

63. Hahamy A, Behrmann M, Malach R (2015): The idiosyncratic brain: distortion of spontaneous connectivity patterns in autism spectrum disorder. *Nat Neurosci* 18: 302–309.

64. Jameel L, Vyas K, Bellesi G, Roberts V, Channon S (2014): Going "Above and Beyond": Are Those High in Autistic Traits Less Pro-social? *J Autism Dev Disord* 44: 1846–1858.

65. Yoshimura Y, Kikuchi M, Ueno S, Okumura E, Hiraishi H, Hasegawa C, *et al.* (2013): The brain's response to the human voice depends on the incidence of autistic traits in the general population. *PloS One* 8: e80126.

66. Kennis M, Rademaker AR, Geuze E (2013): Neural correlates of personality: an integrative review. *Neurosci Biobehav Rev* 37: 73–95.

67. Courchesne E (1997): Brainstem, cerebellar and limbic neuroanatomical abnormalities in autism. *Curr Opin Neurobiol* 7: 269 – 278.

68. Kennedy DP, Adolphs R (2012): The social brain in psychiatric and neurological disorders. *Trends Cogn Sci* 16: 559 – 572.

69. Chevallier C, Kohls G, Troiani V, Brodkin ES, Schultz RT (2012): The social motivation theory of autism. *Trends Cogn Sci* 16: 231 – 239.

70. Zhou Y, Yu F, Duong T (2014): Multiparametric MRI characterization and prediction in autism spectrum disorder using graph theory and machine learning. *PloS One* 9: e90405.

71. Willsey AJ, State MW (2015): Autism spectrum disorders: from genes to neurobiology. *Curr Opin Neurobiol* 30: 92 – 99.

72. Itahashi T, Yamada T, Watanabe H, Nakamura M, Jimbo D, Shioda S, *et al.* (2014): Altered network topologies and hub organization in adults with autism: a resting-state fMRI study. *PloS One* 9: e94115.
73. Marín O (2012): Interneuron dysfunction in psychiatric disorders. *Nat Rev Neurosci* 13: 107–120.

## Supplemental information

ABIDE reproducibility dataset: participants' selection and preprocessing

We selected only ASD participants from the ABIDE database since diagnostic scores were not available for controls. We selected the same 5 datasets used in (1), that is sites with codes: 'CALTECH' (n=19), 'CMU' (n=14), 'PITT' (n=30), 'UM_1' (n=55), 'UM_2' (n=13) for a total of 131 ASD participants. We used version v1.0b of the ABIDE composite phenotypic file to further restrict the ABIDE sample to match our dataset by choosing high functioning male subjects (IQ >=100 for column FIQ, value of 1 for column SEX, a Matlab script for filtering the ABIDE database is available at https://github.com/eglerean/hfASDmodules). Furthermore we required that the subjects had to have valid ADI-R and ADOS scores assessed by professional personnel by filtering subjects with positive values for columns: 'ADI_R_SOCIAL_TOTAL_A', 'ADI_R_VERBAL_TOTAL_BV', 'ADI_RRB_TOTAL_C', 'ADI_R_ONSET_TOTAL_D', 'ADI_R_RSRCH_RELIABLE', 'ADOS_MODULE', 'ADOS_TOTAL', 'ADOS_COMM', 'ADOS_SOCIAL', 'ADOS_STEREO_BEHAV', 'ADOS_RSRCH_RELIABLE'. This yielded 27 subjects with IDs: 51457, 51464, 51468, 51474, 50642, 50643, 50645, 50646, 50647, 50649, 50651, 50652, 50653, 50655, 50002, 50003, 50004, 50006, 50007, 50012, 50016, 50019, 50022, 50024, 50025, 50053, 50056. The ranges (minimum and maximum) for the scores were ADI_R_SOCIAL_TOTAL_A: 10 – 27; ADI_R_VERBAL_TOTAL_BV: 9 – 22; ADI_RRB_TOTAL_C: 2 – 12; ADI_R_ONSET_TOTAL_D: 1 – 5; ADI_R_RSRCH_RELIABLE: 1 – 1; ADOS_TOTAL: 8 – 19; ADOS_COMM: 2 – 6; ADOS_SOCIAL: 5 – 13; ADOS_STEREO_BEHAV: 1 – 6; ADOS_RSRCH_RELIABLE: 1 – 1. We then preprocessed each subject using the same parameters as per our dataset. Since the number of time points was smaller than in our dataset, we applied stricter motion control techniques that is: i) we used a 24 parameters motion regression as explained in (2); ii) To avoid filtering artifacts we discarded the beginning and end of each dataset – see (2); iii) We regressed out signals at ventricles, white matter and cerebral spinal fluid masks as explained in (2); iv) We applied scrubbing so that we kept 125 volumes with lowest framewise displacement for each subject. All subjects had framewise displacement under 0.5mm, except one subject (50003) who had 6 time points above the 0.5 mm threshold (maximum framewise displacement 0.57 mm). We decided to keep this participant anyway since leaving this subject out gave similar results in the final analysis. For the intersubject analysis, we computed normalized Euclidean distance between the joint ADI-R an ADOS scores between each pair of subjects. To further control for head motion during group level analysis, we also computed a intersubject similarity matrix based on mean framewise displacement as a measure of intersubject similarity of average head motion. Head motion similarity did not correlate with ADI-R/ADOS similarity and did not correlate with median scaled inclusivity similarity of subnetworks.

# References

1. Hahamy A, Behrmann M, Malach R (2015): The idiosyncratic brain: distortion of spontaneous connectivity patterns in autism spectrum disorder. *Nat Neurosci* 18: 302–309.

2. Power JD, Mitra A, Laumann TO, Snyder AZ, Schlaggar BL, Petersen SE (2014): Methods to detect, characterize, and remove motion artifact in resting state fMRI. *Neuroimage* 84: 320–341.

Glerean et al. (2015) Reorganization of functional subnetworks in autism

Supplemental tables and figures

| NT module ID | Yeo et al 2011 subnetworks | Correlation (Yeo) | Power et al 2011 subnetworks | Correlation (Power) | Final name given |
|---|---|---|---|---|---|
| 1 | Default | 0.41 | Default mode | 0.42 | Default mode |
| 2 | Default | 0.17 | Default mode | 0.20 | Language |
| 3 | Somatomotor | 0.23 | Auditory | 0.55 | Auditory |
| 4 | Ventral Attention | 0.16 | Salience | 0.63 | Salience |
| 5 | Frontoparietal | 0.13 | Fronto-parietal | 0.26 | Parietal |
| 6 | Dorsal Attention | 0.39 | Dorsal Attention | 0.52 | Dorsal attention |
| 7 | Somatomotor | 0.39 | Sensory/somatomotor Hand | 0.60 | Sensorimotor |
| 8 | Visual | 0.33 | Visual | 0.50 | Visual primary |
| 9 | Limbic | 0.14 | Subcortical | 0.75 | Ventro-temporal limbic |
| 10 | Default | 0.05 | Memory retrieval | 0.45 | Precuneus |
| 11 | Visual | 0.03 | Cerebellar | 0.73 | Cerebellum |
| 12 | Visual | 0.43 | Visual | 0.62 | Visual extrastriate |

Table S1 – This table shows the highest amount of overlap between each NT group consensus subnetwork and previously defined subnetworks in the literature. Overlap is measured with spatial Pearson's correlation. The peak value between the two reference subnetworks was used in choosing the final name of each component. Subnetworks for Yeo et al 2011: Visual, Somatomotor, Dorsal attention, Ventral attention, Limbic, Frontoparietal, Default (Cerebellum and sub-cortical areas were not included in Yeo et al. 2011). Subnetworks for Power et al. 2011: Sensory/somatomotor Hand, Sensory/somatomotor Mouth, Cingulo-opercular Task Control, Auditory, Default-mode, Memory retrieval, Visual, Fronto-parietal Task Control, Salience, Subcortical, Ventral attention, Dorsal attention, Cerebellar. Final labels were chosen manually, after considering the overlap with the two previous studies. Code to compute correlation of subnetworks is available at https://github.com/eglerean/hfASDmodules/compare_modules

Glerean et al. (2015) Reorganization of functional subnetworks in autism

| Anatomical name | AAL (cerebral cortex) & HO (sub-cortex and cerebellum) labels | Major region label | Acronym (as per doi:10.1371/ journal.pone.0005226) |
|---|---|---|---|
| Precentral gyrus (L) | Precentral_L | Frontal | PreCG(L) |
| Precentral gyrus (R) | Precentral_R | Frontal | PreCG(R) |
| Superior frontal gyrus (L) | Frontal_Sup_L | Frontal | SFGdor(L) |
| Superior frontal gyrus (R) | Frontal_Sup_R | Frontal | SFGdor(R) |
| Orbital superior frontal gyrus (L) | Frontal_Sup_Orb_L | Frontal | ORBsup(L) |
| Orbital superior frontal gyrus (R) | Frontal_Sup_Orb_R | Frontal | ORBsup(R) |
| Middle frontal gyrus (L) | Frontal_Mid_L | Frontal | MFG(L) |
| Middle frontal gyrus (R) | Frontal_Mid_R | Frontal | MFG(R) |
| Orbital middle frontal gyrus (L) | Frontal_Mid_Orb_L | Frontal | ORBmid(L) |
| Orbital middle frontal gyrus (R) | Frontal_Mid_Orb_R | Frontal | ORBmid(R) |
| Opercular inferior frontal gyrus (L) | Frontal_Inf_Oper_L | Frontal | IFGoperc(L) |
| Opercular inferior frontal gyrus (R) | Frontal_Inf_Oper_R | Frontal | IFGoperc(R) |
| Triangular inferior frontal gyrus (L) | Frontal_Inf_Tri_L | Frontal | IFGtriang(L) |
| Triangular inferior frontal gyrus (R) | Frontal_Inf_Tri_R | Frontal | IFGtriang(R) |
| Orbital inferior frontal gyrus (L) | Frontal_Inf_Orb_L | Frontal | ORBinf(L) |
| Orbital inferior frontal gyrus (R) | Frontal_Inf_Orb_R | Frontal | ORBinf(R) |
| Rolandic operculum (L) | Rolandic_Oper_L | Frontal | ROL(L) |
| Rolandic operculum (R) | Rolandic_Oper_R | Frontal | ROL(R) |
| Supplementary motor area (L) | Supp_Motor_Area_L | Frontal | SMA(L) |
| Supplementary motor area (R) | Supp_Motor_Area_R | Frontal | SMA(R) |
| Olfactory cortex (L) | Olfactory_L | Frontal | OLF(L) |
| Olfactory cortex (R) | Olfactory_R | Frontal | OLF(R) |
| Medial superior frontal gyrus (L) | Frontal_Sup_Medial_L | Frontal | SFGmed(L) |
| Medial superior frontal gyrus (R) | Frontal_Sup_Medial_R | Frontal | SFGmed(R) |
| Orbital medial frontal gyrus (L) | Frontal_Mid_Orb_L | Frontal | ORBsupmed(L) |
| Orbital medial frontal gyrus (R) | Frontal_Mid_Orb_R | Frontal | ORBsupmed(R) |
| Gyrus rectus (L) | Rectus_L | Frontal | REC(L) |
| Gyrus rectus (R) | Rectus_R | Frontal | REC(R) |
| Insula (L) | Insula_L | Insula | INS(L) |
| Insula (R) | Insula_R | Insula | INS(R) |
| Anterior cingulum (L) | Cingulum_Ant_L | Cingulate | ACG(L) |
| Anterior cingulum (R) | Cingulum_Ant_R | Cingulate | ACG(R) |
| Middle cingulum (L) | Cingulum_Mid_L | Cingulate | DCG(L) |
| Middle cingulum (R) | Cingulum_Mid_R | Cingulate | DCG(R) |
| Posterior cingulum (L) | Cingulum_Post_L | Cingulate | PCG(L) |
| Posterior cingulum (R) | Cingulum_Post_R | Cingulate | PCG(R) |
| Parahippocampal gyrus (L) | ParaHippocampal_L | Occipital | PHG(L) |
| Parahippocampal gyrus (R) | ParaHippocampal_R | Occipital | PHG(R) |
| Calcarine gyrus (L) | Calcarine_L | Occipital | CAL(L) |
| Calcarine gyrus (R) | Calcarine_R | Occipital | CAL(R) |
| Cuneus (L) | Cuneus_L | Occipital | CUN(L) |
| Cuneus (R) | Cuneus_R | Occipital | CUN(R) |
| Lingual gyrus (L) | Lingual_L | Occipital | LING(L) |
| Lingual gyrus (R) | Lingual_R | Occipital | LING(R) |
| Superior occipital gyrus (L) | Occipital_Sup_L | Occipital | SOG(L) |
| Superior occipital gyrus (R) | Occipital_Sup_R | Occipital | SOG(R) |
| Middle occipital gyrus (L) | Occipital_Mid_L | Occipital | MOG(L) |
| Middle occipital gyrus (R) | Occipital_Mid_R | Occipital | MOG(R) |
| Inferior occipital gyrus (L) | Occipital_Inf_L | Occipital | IOG(L) |
| Inferior occipital gyrus (R) | Occipital_Inf_R | Occipital | IOG(R) |
| Fusiform gyrus (L) | Fusiform_L | Occipital | FFG(L) |

Glerean et al. (2015) Reorganization of functional subnetworks in autism

| Fusiform gyrus (R) | Fusiform_R | Occipital | FFG(R) |
|---|---|---|---|
| Postcentral gyrus (L) | Postcentral_L | Parietal | PoCG(L) |
| Postcentral gyrus (R) | Postcentral_R | Parietal | PoCG(R) |
| Superior parietal lobule (L) | Parietal_Sup_L | Parietal | SPG(L) |
| Superior parietal lobule (R) | Parietal_Sup_R | Parietal | SPG(R) |
| Inferior parietal lobule (L) | Parietal_Inf_L | Parietal | IPL(L) |
| Inferior parietal lobule (R) | Parietal_Inf_R | Parietal | IPL(R) |
| Supramarginal gyrus (L) | SupraMarginal_L | Parietal | SMG(L) |
| Supramarginal gyrus (R) | SupraMarginal_R | Parietal | SMG(R) |
| Angular gyrus (L) | Angular_L | Parietal | ANG(L) |
| Angular gyrus (R) | Angular_R | Parietal | ANG(R) |
| Precuneus (L) | Precuneus_L | Parietal | PCUN(L) |
| Precuneus (R) | Precuneus_R | Parietal | PCUN(R) |
| Paracentral lobule (L) | Paracentral_Lobule_L | Parietal | PCL(L) |
| Paracentra lobule (R) | Paracentral_Lobule_R | Parietal | PCL(R) |
| Heschl gyrus (L) | Heschl_L | Temporal | HES(L) |
| Heschl gyrus (R) | Heschl_R | Temporal | HES(R) |
| Superior temporal gyrus (L) | Temporal_Sup_L | Temporal | STG(L) |
| Superior temporal gyrus (R) | Temporal_Sup_R | Temporal | STG(R) |
| Temporal pole (superior) (L) | Temporal_Pole_Sup_L | Temporal | TPOsup(L) |
| Temporal pole (superior) (R) | Temporal_Pole_Sup_R | Temporal | TPOsup(R) |
| Middle temporal gyrus (L) | Temporal_Mid_L | Temporal | MTG(L) |
| Middle temporal gyrus (R) | Temporal_Mid_R | Temporal | MTG(R) |
| Temporal pole (middle) (L) | Temporal_Pole_Mid_L | Temporal | TPOmid(L) |
| Temporal pole (middle) (R) | Temporal_Pole_Mid_R | Temporal | TPOmid(R) |
| Inferior temporal gyrus (L) | Temporal_Inf_L | Temporal | ITG(L) |
| Inferior temporal gyrus (R) | Temporal_Inf_R | Temporal | ITG(R) |
| Thalamus (L) | Left_Thalamus | Subcortex | THA(L) |
| Caudate (L) | Left_Caudate | Subcortex | CAU(L) |
| Putamen (L) | Left_Putamen | Subcortex | PUT(L) |
| Pallidum (L) | Left_Pallidum | Subcortex | PAL(L) |
| Brainstem | Brain-Stem | Subcortex | BST |
| Hippocampus (L) | Left_Hippocampus | Subcortex | HIP(L) |
| Amygdala (L) | Left_Amygdala | Subcortex | AMYG(L) |
| Nucleus accumbens (L) | Left_Accumbens | Subcortex | NAcc(L) |
| Thalamus (R) | Right_Thalamus | Subcortex | THA(R) |
| Caudate (R) | Right_Caudate | Subcortex | CAU(R) |
| Putamen (R) | Right_Putamen | Subcortex | PUT(R) |
| Pallidum (R) | Right_Pallidum | Subcortex | PAL(R) |
| Hippocampus (R) | Right_Hippocampus | Subcortex | HIP(R) |
| Amygdala (R) | Right_Amygdala | Subcortex | AMYG(R) |
| Nucleus accumbens (R) | Right_Accumbens | Subcortex | NAcc(R) |
| Cerebellar lobule I-IV (L) | Left_I-IV | Cerebellum | I-IV(L) |
| Cerebellar lobule I-IV (R) | Right_I-IV | Cerebellum | I-IV(R) |
| Cerebellar lobule V (L) | Left_V | Cerebellum | V(L) |
| Cerebellar lobule V (R) | Right_V | Cerebellum | V(R) |
| Cerebellar lobule VI (L) | Left_VI | Cerebellum | VI(L) |
| Cerebellar vermis VI | Vermis_VI | Cerebellum | VI-vermis |
| Cerebellar lobule VI (R) | Right_VI | Cerebellum | VI(R) |
| Cerebellar crus I (L) | Left_Crus_I | Cerebellum | XI(L) |
| Cerebellar crus I (R) | Right_Crus_I | Cerebellum | XI(R) |
| Cerebellar crus II (L) | Left_Crus_II | Cerebellum | XII(L) |
| Cerebellar vermis crus II | Vermis_Crus_II | Cerebellum | XII-vermis |
| Cerebellar crus II (R) | Right_Crus_II | Cerebellum | XII(R) |
| Cerebellar lobule VIIb (L) | Left_VIIb | Cerebellum | VIIb(L) |

| Cerebellar lobule VIIb (R) | Right_VIIb | Cerebellum | VIIb(R) |
| Cerebellar lobule VIIIa (L) | Left_VIIIa | Cerebellum | VIIIa(L) |
| Cerebellar vermis VIIIa | Vermis_VIIIa | Cerebellum | VIIIa-vermis |
| Cerebellar lobule VIIIa (R) | Right_VIIIa | Cerebellum | VIIIa(R) |
| Cerebellar lobule VIIIb (L) | Left_VIIIb | Cerebellum | VIIIb(L) |
| Cerebellar vermis VIIIb | Vermis_VIIIb | Cerebellum | VIIIb-vermis |
| Cerebellar lobule VIIIb (R) | Right_VIIIb | Cerebellum | VIIIb(R) |
| Cerebellar lobule IX (L) | Left_IX | Cerebellum | IX(L) |
| Cerebellar vermis IX | Vermis_IX | Cerebellum | IX-vermis |
| Cerebellar lobule IX (R) | Right_IX | Cerebellum | IX(R) |
| Cerebellar lobule X (L) | Left_X | Cerebellum | X(L) |
| Cerebellar vermis X | Vermis_X | Cerebellum | X-vermis |
| Cerebellar lobule X (R) | Right_X | Cerebellum | X(R) |

Table S2 – In this table we report the list of acronyms used for visualization and automatic labeling of the modules. The acronyms were the same as in He, Y., Wang, J., Wang, L., Chen, Z. J., Yan, C., Yang, H., … Evans, A. C. (2009). Uncovering intrinsic modular organization of spontaneous brain activity in humans. *PloS One*, *4*(4), e5226. doi:10.1371/journal.pone.0005226

.

Glerean et al. (2015) Reorganization of functional subnetworks in autism

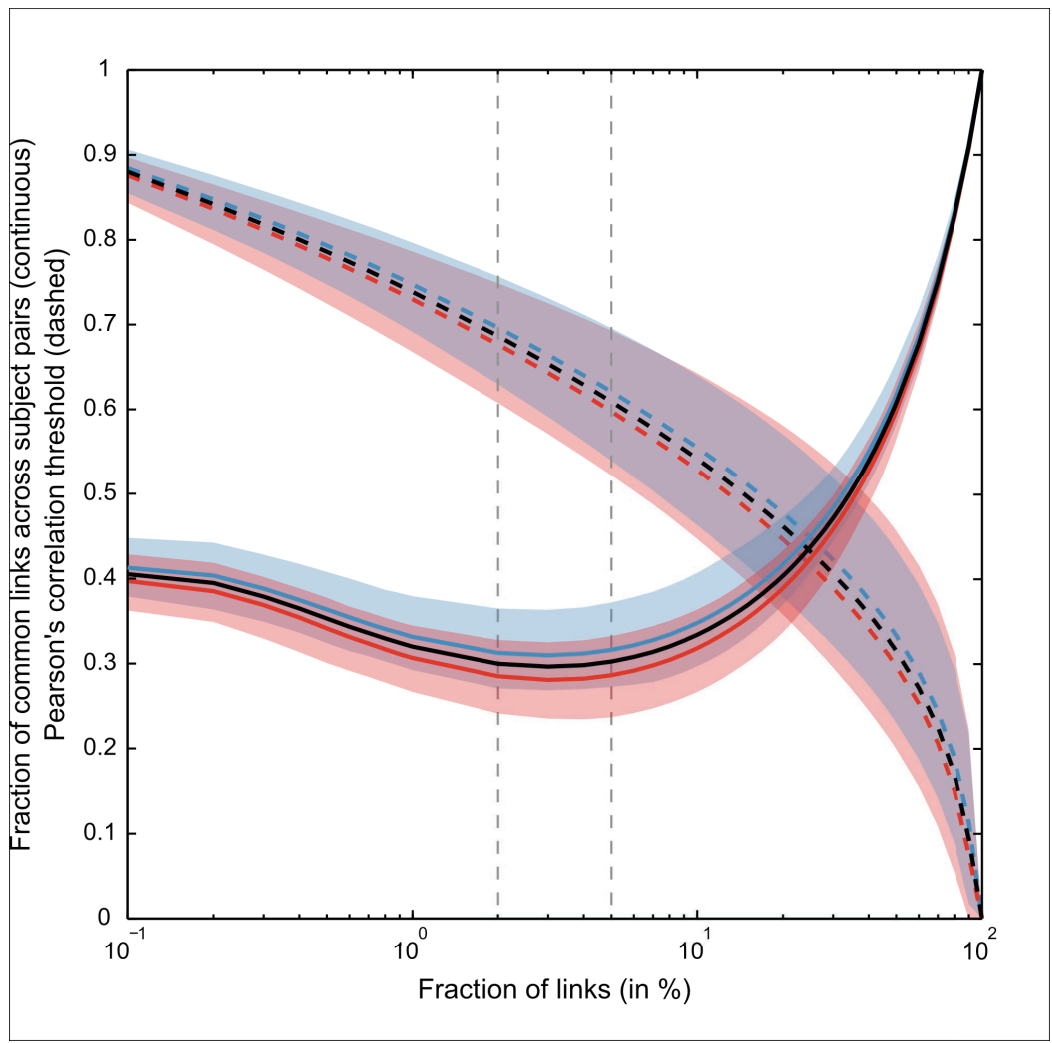

Figure S1 – We thresholded the networks by link density such that in each network, the fraction of links with highest correlations corresponding to a given link density was retained. For each density value (from 0.1 to 100%, with step 0.1 for densities < 1% and step 1 for densities > 1%) we calculated the link overlap measured in fraction of links (black continuous line), the overlap within NT participants (blue continuous line, 5% and 95% confidence area in transparent blue) and the overlap within ASD participants (red continuous line, 5% and 95% confidence area in transparent red). We also show in the same plot the values of average correlation thresholds corresponding to each link density (i.e. the correlation coefficient corresponding to the last retained link). These are shown across all subjects (dashed black line), across NT (dashed blue line, 5% and 95% confidence area in transparent blue) and across ASD (dashed red line, 5% and 95% confidence area in transparent red). When the density approaches 100%, the correlation threshold goes to zero and the overlap across participants is maximal since all links are retained in the network. For small densities, the maximum of the overlap across participants can be interpreted in terms of network backbones. The overlap across individual networks was least for densities between 2% (equivalent to a threshold of r = ~0.69 for the average subject) and 5% (equivalent to a threshold of r = ~0.61).

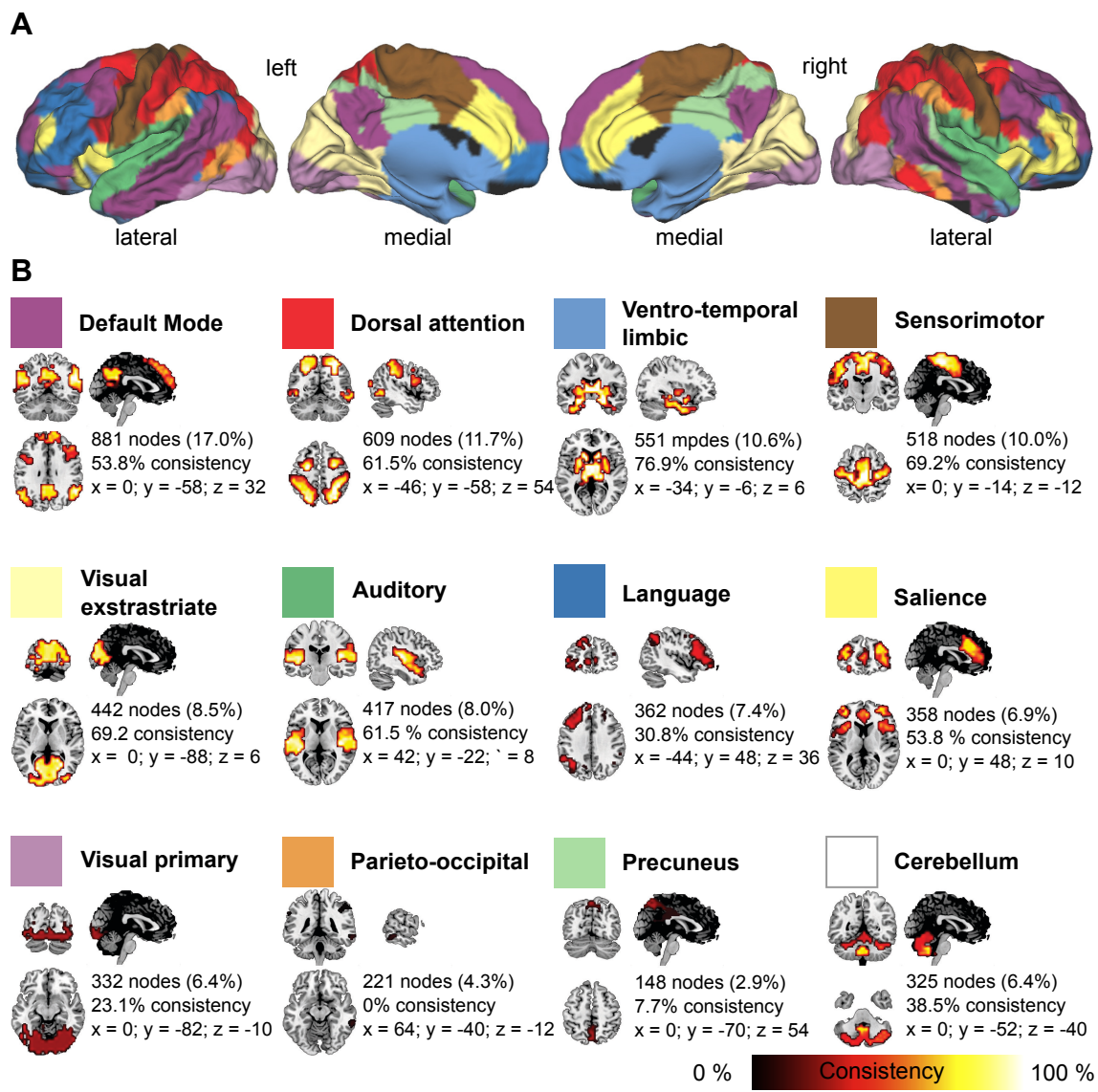

Figure S2 – A) Group consensus subnetworks for the control participants. Each subnetwork was color coded and the color code is consistent with Figure 2 in the manuscript. B) Each group consensus subnetwork is plotted separately by showing the level of consistency across brain areas. Median consistency and number of nodes are reported next to each subnetwork. Labelling of subnetwork was done quantitatively based on published subnetworks atlases, see Table S1. Full browsable maps for NT and ASD consensus modules are available at http://neurovault.org/collections/437/.

Glerean et al. (2015) Reorganization of functional subnetworks in autism

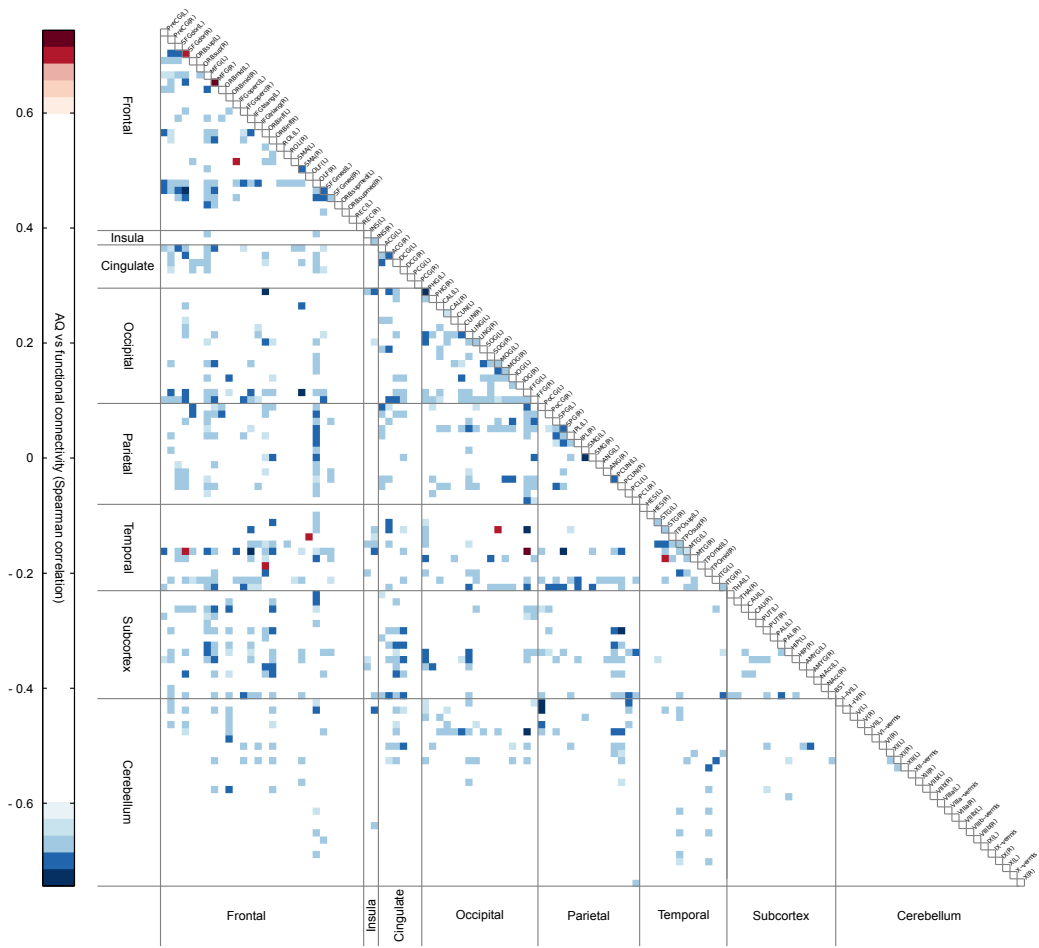

Figure S3 – Summary connectivity matrix. The full connectivity matrix (5183x5183) was downsampled for visualization purposes so that nodes belonging to the same anatomical regions – as defined by the AAL atlas – were grouped together and non-null links were averaged. The legend for this figure is the same of figure 4B, where 4B is composed of a subset of regions extracted from this figure. The full list of abbreviations is reported in Table S2.

Glerean et al. (2015) Reorganization of functional subnetworks in autism

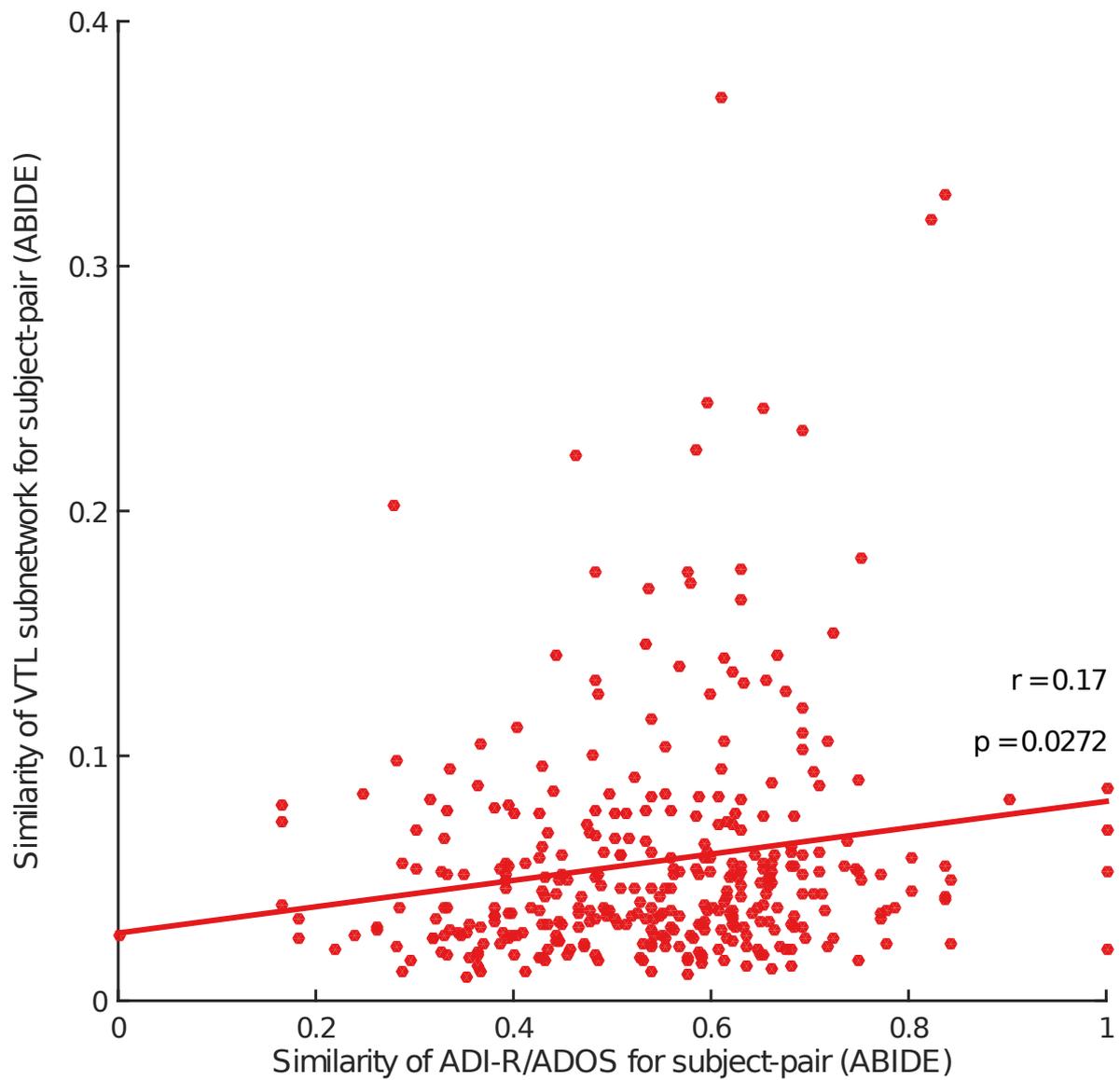

Figure S4 – Mantel test showing association between VTL subnetwork structure and autistic symptoms in the ABIDE replication dataset. Each dot is a pair of ASD subjects (351 pairs for 27 subjects) showing their VTL subnetwork similarity with median scaled inclusivity and behavioral similarity with ADI-R and ADOS score vectors. Effect size is reported as correlation value and p-value was computed with permutations.